\documentclass[sigconf]{acmart}

\usepackage[most, breakable]{tcolorbox}  
\usepackage[utf8]{inputenc}

\usepackage{newtxmath} 
\usepackage{array}
\usepackage{multirow}
\usepackage{graphicx}
\usepackage[ruled,vlined,lined,commentsnumbered]{algorithm2e}
\usepackage{balance}
\usepackage{algorithmic}
\usepackage{esvect}
\usepackage{booktabs}
\usepackage{tabularx}
\usepackage{anyfontsize}
\usepackage[all]{nowidow}
\usepackage[normalem]{ulem}
\usepackage{xcolor}
\usepackage{colortbl}
\usepackage{enumitem}
\usepackage{makecell}
\usepackage{caption}
\usepackage{subcaption}
\usepackage{framed}
\usepackage{wrapfig}

\usepackage{amsmath,amssymb}
\usepackage{natbib}
\usepackage{float}
\usepackage{lineno}
\usepackage{soul}
\usepackage{flushend}
\usepackage{multicol}

\newcolumntype{Y}{>{\centering\arraybackslash}X}

\definecolor{commentGray}{RGB}{120,120,120}
\renewcommand{\algorithmiccomment}[1]{\bgroup\color{commentGray}{//#1}\egroup}

\usepackage{pifont}
\usepackage{listings}
\usepackage{framed}
\usepackage{tikz}
\definecolor{light-gray}{gray}{0.9}
\usepackage[framemethod=TikZ]{mdframed}
\mdfsetup{skipabove=5pt,skipbelow=3pt}
\usepackage{lipsum}
\mdfdefinestyle{RQFrame}{%
	linecolor=black,
	outerlinewidth=0.15pt,
	roundcorner=3pt,
	innertopmargin=2pt,
	innerbottommargin=2pt,
	innerrightmargin=4pt,
	innerleftmargin=4pt,
	backgroundcolor=light-gray}

\newcommand{\Lagr}{\mathcal{L}}

\definecolor{javagreen}{rgb}{0.25,0.5,0.35} 

\lstdefinestyle{Alg}{
  basicstyle=\ttfamily\footnotesize,
  breaklines=true,
  tabsize=2,
  mathescape,
  numbers=left,
  xleftmargin=2.5em,
  xrightmargin=0.5em,
  frame=tb,
  framexleftmargin=2em,
  emph={Algorithm,Input,Output,for,each,do,if,else,Function,while,let,be,repeat,until,return,times,and,or,break,in,then,},
  emphstyle={\textbf},
  escapechar=?,
  morecomment=[l][\color{javagreen}]{//},
  columns=flexible,
}

\newcounter{commentnumber}

\begin{document}



%
\title[Bridging the Gap between Real-world and Synthetic Images for Testing ADS]{Bridging the Gap between Real-world and Synthetic Images for Testing Autonomous Driving Systems}
%
%
%


\author{Mohammad Hossein Amini and  Shiva Nejati}
\affiliation{%
  \institution{University of Ottawa, Ottawa, Canada}
  \country{}
}
\email{{mh.amini,snejati}@uottawa.ca}

\begin{abstract}

Deep Neural Networks (DNNs) for Autonomous Driving Systems (ADS) are typically trained on real-world images and tested using synthetic images from simulators. This approach  results in training and test datasets with dissimilar distributions, which can potentially  lead to erroneously decreased test accuracy. To address this issue, the  literature suggests applying domain-to-domain translators to test datasets to bring them closer to the training datasets. However, translating images used for testing may unpredictably affect the reliability, effectiveness and efficiency of the testing process.  Hence, this paper investigates the following questions in the context of ADS: \emph{Could translators reduce the effectiveness of  images used for ADS-DNN testing and their ability to reveal faults in ADS-DNNs? Can translators result in excessive time overhead during simulation-based testing?} To address these questions, we consider three domain-to-domain translators: CycleGAN and neural style transfer, from the  literature, and SAEVAE, our proposed translator.  Our results for two critical ADS tasks -- lane keeping and object detection -- indicate that translators significantly narrow the gap in ADS test accuracy caused by distribution dissimilarities between training and test data, with SAEVAE outperforming the other two translators. We  show that, based on the recent diversity, coverage, and fault-revealing ability metrics for testing deep-learning systems, translators do \emph{not} compromise the diversity and the coverage of test data  \emph{nor} do they lead to revealing fewer faults in ADS-DNNs. Further, among the  translators considered, SAEVAE incurs a negligible overhead in simulation time and can be efficiently integrated into simulation-based testing. Finally, we show that translators increase the correlation between offline and simulation-based testing results, which can help reduce the cost of simulation-based testing. Our  \textbf{replication package} is available online~\cite{github}.

\end{abstract}

\maketitle

\section{Introduction}
\label{sec:intro}
Deep Neural Networks (DNNs) for Autonomous Driving Systems (ADS) typically undergo two main types of rigorous testing~\cite{zhang_ml_testing}: (1)~\emph{Offline} testing~\cite{system_level}, which evaluates DNNs as standalone units using static datasets. (2)~\emph{Online} or simulation-based testing~\cite{lei,AminiNN24,MatinnejadNBBP13,GonzalezVNBI18,NejatiSSFMM23}, which involves embedding DNNs into simulators to test them in interaction with their environment. Both offline and online testing are essential for ensuring the safety and reliability of ADS-DNNs.

\begin{figure}[t]
     \centering
     \begin{subfigure}[b]{0.30\columnwidth}
         \centering
         \includegraphics[width=\textwidth]{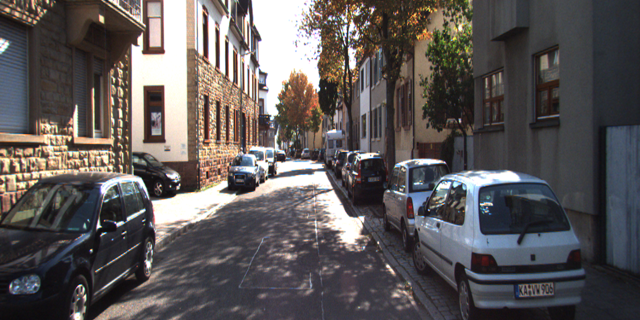}
         \caption{Real-world image}
     \end{subfigure}
     \hfill
     \begin{subfigure}[b]{0.30\columnwidth}
         \centering
         \includegraphics[width=\textwidth]{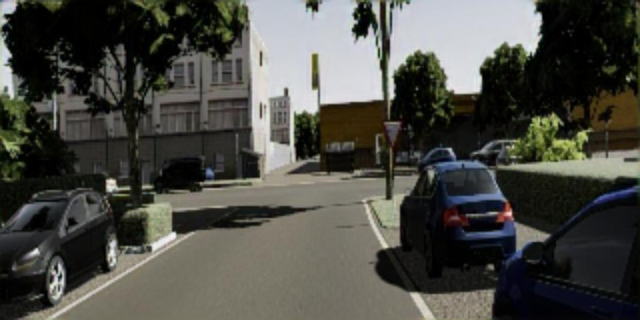}
         \caption{Synthetic image}
     \end{subfigure}
     \hfill
     \begin{subfigure}[b]{0.30\columnwidth}
         \centering
         \includegraphics[width=\textwidth]{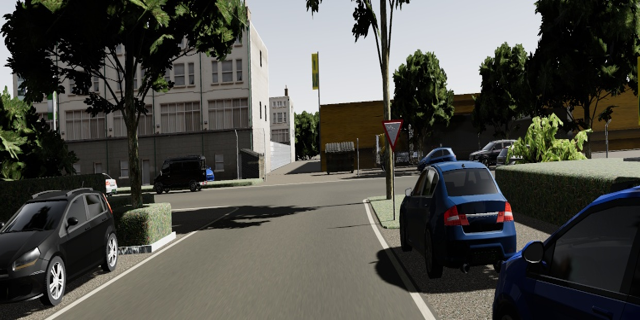}
         \caption{Translated image}
     \end{subfigure}
     \vspace*{-0.25cm}
        \caption{Three sample images for ADS: (a) an image from a real-world dataset used to train an ADS-DNN; (b) a simulator-generated image for ADS-DNN testing; and (c) the transformation of image (b) using the SAEVAE translator}
        \label{fig:samples}
        \vspace*{0.1cm}
\end{figure}

Typically, we use real-world images to train ADS-DNNs. But, for testing, particularly since online testing is required, images from simulators are inevitably used. Although advanced simulators can produce highly realistic images, there are still noticeable differences in style and texture compared to real-world images. For example, Figure~\ref{fig:samples}(a) shows a real-world image related to the ADS object-detection task, whereas Figure~\ref{fig:samples}(b) presents a synthetic image for the same task, but with dissimilarities to the image in Figure~\ref{fig:samples}(a).  Due to these dissimilarities,  we may test ADS-DNNs on images that have a different distribution than those used for training them. However, it is recommended to test a DNN using a dataset that is close to the distribution of its training dataset~\cite{goodfellow}. Otherwise, neglecting this could lead to erroneous inaccuracies in test results.

Several deep-learning domain-to-domain translator models have been proposed to convert an image from one data distribution to a different, but related, data distribution. For example, Pix2Pix~\cite{pix2pix} and CycleGAN~\cite{cyclegan} use generative adversarial networks (GANs) to adapt an image from one distribution to  a corresponding image in another distribution, e.g., changing an image from a winter scene to its summer counterpart. In contrast, neural style transfer methods~\cite{gatys, deep_photo_style, style_review} take an image and modify it to embody a desired style, e.g., texture, while preserving the image's primary content. 

Recently, CycleGANs have been used to address the distribution gap between the training and testing datasets for ADS-DNNs~\cite{mind_the_gap, two_is_better, stoccoicst}. Specifically, instead of using synthetic images directly for offline or online ADS testing, these images are first converted using CycleGANs and then fed to ADS-DNNs. This approach reduces the likelihood of ADS-DNNs encountering images outside their training set distribution during testing, resulting in a more realistic evaluation of ADS-DNNs. While these recent studies~\cite{mind_the_gap, two_is_better, stoccoicst} and the adoption of translators in the deep-learning community suggest that translators may be necessary for ADS testing,  the following key questions related to the reliability, effectiveness, and efficiency of translators for ADS testing remain unanswered:

\begin{tcolorbox}[colback=gray!10!white,colframe=black!75!black, breakable]
Can translators compromise the quality of test data in terms of diversity, coverage, or fault-revealing ability?  Can synthetic images, before applying translators, reveal more flaws in an ADS-DNN than the translated versions of the synthetic images? Are translators efficient enough for online testing? Can translators help reduce the overall effort needed for ADS testing?
\end{tcolorbox}

This paper presents an extensive study to answer the above questions. We compare the accuracy of ADS-DNNs and the quality of the test data for ADS-DNN testing before and after applying translators for both offline and online testing. In addition to the offline and online accuracy metrics, we use test data quality metrics including white-box test coverage metrics, specifically neuron coverage~\cite{neuron_coverage} and surprise adequacy~\cite{surprise}, and black-box test data diversity metrics, i.e.,  geometric diversity~\cite{geometric} and standard deviation~\cite{clustering}. Further, we evaluate the fault-revealing ability of test data for ADS-DNNs using a clustering-based fault \hbox{estimation metric~\cite{clustering}.} 

As for the ADS-DNNs under test, we use five public-domain DNNs automating two critical ADS tasks: lane keeping and object detection. We use two widely-used, public datasets, the Udacity Jungle ~\cite{udacity_jungle} and KITTI~\cite{kitti} for ADS-DNN training. For KITTI, a dataset representing its synthetic counterpart, known as  vKITTI~\cite{vkitti} is provided for testing. For  Udacity Jungle, we develop a synthetic counterpart using the ADS simulator, BeamNG~\cite{beamng, sbftgithub}, and use the same BeamNG simulator for online testing. We use three image-to-image translators that represent alternative architectures within this category of deep learning models: CycleGAN~\cite{cyclegan}, which is a generative  translator model, neural style transfer~\cite{gatys}, which is an image perturbation method, and  SAEVAE, our proposed translator architecture which aims to address the limitations of CycleGAN and neural style transfer. Specifically, SAEVAE aims to mitigate the high cost of training for CycleGAN and high translation time of neural style transfer.

Our study starts by confirming the presence of a distribution gap between the real-world and synthetic datasets used for training and testing ADS-DNNs, and demonstrating the effectiveness of translators in bringing the data distributions closer and improving ADS-DNNs testing results in online and offline tests. The effectiveness results for translators  are established through the following three research questions addressed in our paper:

\textbf{RQ1.} \emph{How well do translators mitigate the distribution gap between training  and test datasets?} \textbf{Answer.}  Our assessment, which is  independent from the ADS-DNNs under test,  shows that translators can effectively align the  training and test datasets. 

\textbf{RQ2.} \emph{How well do translators mitigate the erroneous accuracy decline caused by disparities between training and test datasets in offline testing results?} \textbf{Answer.} Domain-to-domain translators can significantly reduce the accuracy decline in offline testing, achieving up to a 57\%  accuracy 
 improvement  for the lane-keeping task and up to a 19\%  accuracy improvement for the object-detection task. 

\textbf{RQ3.} \emph{How well do translators reduce the occurrences and severity of failures in online ADS testing?} \textbf{Answer.} Translators can reduce the frequency and severity of failures by up to 80\% in online testing.

We then answer the following questions to evaluate the reliability and efficiency of translators in ADS testing. Specifically, the questions assess the impact of translators on the quality of test data, the time overhead incurred by using translators for online testing, and the potential effort savings in ADS testing  using translators.

\textbf{RQ4.} \emph{Can translators preserve the quality of test data in terms of diversity, coverage and fault-revealing ability?} \textbf{Answer.} Based on five test quality metrics for DNN testing, translators in most cases preserve the quality of test data, and only slightly reduce test data quality in a few cases.

\textbf{RQ5.} \emph{Is the time overhead incurred by using translators during online testing justified?} \textbf{Answer:} Among the three studied translators, SAEVAE has the lowest time overhead, incurring only an 8\% increase in simulation time during online testing.

\textbf{RQ6.} \emph{Can translators reduce the cost of ADS testing by increasing the correlation between offline and online testing results?} \textbf{Answer.}  Translators improve the correlation between offline and online test results for three of our four ADS-DNNs. Increasing this correlation may help reduce the need to online testing, which is expensive, and replace it with more offline testing, which is less expensive.

\vspace*{.2cm}
\textbf{Novelty.}  We present the \emph{first} study to assess the effectiveness, reliability and efficiency of image-to-image translators for ADS testing. Our empirical study is extensive and encompasses  three different image-to-image translation methods for online and offline testing of ADS lane-keeping and object-detection tasks, using five ADS-DNNs, two public-domain datasets and the BeamNG simulator. We assess the reliability of translators in preserving test data quality with respect to five widely-used metrics~\cite{neuron_coverage, surprise, geometric, clustering, clustering:GIST, clustering}.

\vspace*{.2cm}
\textbf{Relevance.} The use of synthetic and simulator-generated images for testing DNN-enabled systems is inevitable in many domains, including ADS. Image-to-image translators are key to mitigating the erroneous accuracy results for DNN-enabled systems caused by the misalignment of test and training datasets, which is a by-product of using synthetic images for testing. Our paper studies the effectiveness, reliability, and efficiency of image-to-image translators for ADS testing. Our results are important to software engineering since we cannot reliably integrate translators in our testing practices without measuring their impact on our testing practices.

\section{Image-to-Image Translation}
\label{sec:saevae}
In this section,  we review the two widely-used image-to-image translator models, GAN-based and neural style, and outline their limitations for ADS testing~\cite{cyclegan, pix2pix, gatys, style_review, deep_photo_style}. We then introduce an alternative translator, SAEVAE, proposed in this paper. SAEVAE aims to mitigate the high cost of training for CycleGAN and high translation time of neural style transfer.

\textbf{GAN-based Translators.} GAN-based translators require as input two datasets from two different domains. Some GAN-based methods, e.g.,  Pix2Pix~\cite{pix2pix}, require paired datasets, which demand extra manual effort. Thus,  we consider CycleGAN~\cite{cyclegan} from the family of GAN-based translators that does not require dataset pairing. CycleGAN utilizes two GANs, each composed of a generator and a discriminator. One GAN learns to translate images from domain A to domain B, and the other learns the reverse. Each generator's job is to produce images that look as close as possible to those in its target domain, while each discriminator's job is to distinguish between images from its source and target domains. 

\emph{Limitation of CycleGAN:} Training CycleGAN models is complicated and time-consuming because the simultaneous training of discriminator and generator models makes it difficult to reach equilibrium during the training process~\cite{gan_unstable_salimans, gan_unstable_how_good}. 

\textbf{Neural Style Translators.} Neural style transfer methods, unlike GAN-based translators, only need two images: one for style and another for content. The style image provides patterns and textures, while the content image supplies the objects. This technique modifies the style of the content image to match the style image, while retaining the original content. A notable technique by Gatys et al. \cite{gatys} uses a pretrained convolutional neural network (CNN) \cite{vgg19} to separate style and content from images. Early CNN layers capture the image's style, while deeper layers focus on content. The process involves fixing the style image and the pretrained CNN iteratively modifying the content image to match the style by minimizing a loss function.

\emph{Limitation of neural style translators:}  Neural style translators iteratively adjust the content image using gradient descent,  leading to a high execution time.

\textbf{SAEVAE Translators.} We introduce SAEVAE, a novel translator that aims to address the high training  cost  for CycleGANs and the high translation time for neural style transfer models. SAEVAE consists of two sequential components:  (1)~a Sparse Auto Encoder (SAE)~\cite{sae}, and (2)~a Variational Auto Encoder (VAE)~\cite{vae}, hence the acronym SAEVAE. Both SAE and VAE are auto-encoders (AE), i.e., models that reproduce their inputs as outputs.

Let $\mathit{D}_{\mathit{real}}$ and $\mathit{D}_{\mathit{sim}}$ be the real-world and the synthetic datasets, respectively. SAEVAE converts any image in $\mathit{D}_{\mathit{sim}}$ into a corresponding image in the distribution of $\mathit{D}_{\mathit{real}}$. The SAE's role  is to extract and reconstruct the most influential features of $\mathit{D}_{\mathit{sim}}$, while the VAE's role  is to validate how closely the image generated by the SAE matches the distribution of $\mathit{D}_{\mathit{real}}$.

Figure \ref{fig:saevae_overview} shows the training process of SAEVAE translators. We denote the SAE model by $F_S(x)$, and  the VAE model  by $F_V(x)$. First, the VAE is trained on $\mathit{D}_{\mathit{real}}$ so that it  distinguishes samples belonging to $\mathit{D}_{\mathit{real}}$'s distribution, as these samples exhibit a lower reconstruction error compared to those from other distributions.  Once the training of the VAE is complete, we freeze its parameters, meaning that no further VAE training occurs in subsequent steps. We denote the frozen VAE by $F^*_V(x)$.  Next, we train the SAE on $\mathit{D}_{\mathit{sim}}$ using the loss function $\Lagr$ defined as follows:  

\begin{figure}[t]
    \centering
    \includegraphics[width=0.7\columnwidth]{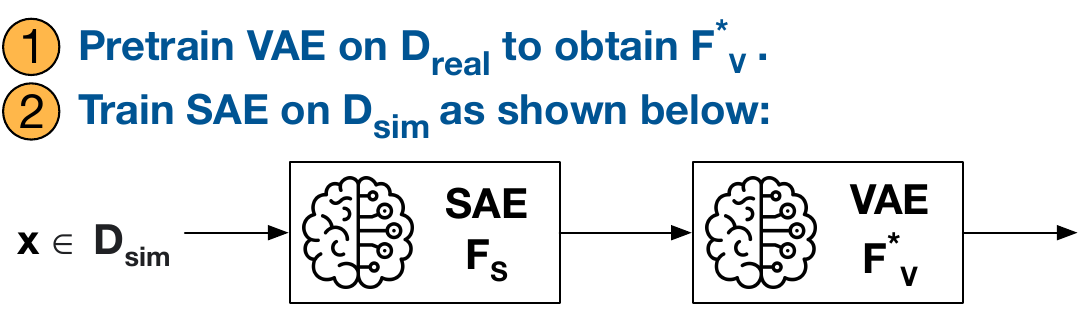}
    \vspace*{-.2cm}
    \caption{The training process for our SAEVAE translators}
    \label{fig:saevae_overview}
    \vspace*{.05cm}
\end{figure}

{\small\begin{align}
 \Lagr &= \Lagr_{\mathit{SAE}} + \alpha\Lagr_{\mathit{VAE}} \label{eqn:saevae_loss}\\
    \Lagr_{\mathit{SAE}} &= ||F_S(x) - x||_1\\
    \Lagr_{\mathit{VAE}} &= ||F^*_V(F_S(x)) - F_S(x)||_1 \label{fm:lvae}
\end{align}}

where $x \in \mathit{D}_{\mathit{sim}}$, $||\cdot||_1$ indicates the $l_1$ distance, and 
$\alpha$ is a positive hyperparameter that tunes the balance between adherence to the  distribution of $\mathit{D}_{\mathit{real}}$ versus  the preservation of the content of the input image from $\mathit{D}_{\mathit{sim}}$.   The $\Lagr_{\mathit{SAE}}$ term aims to reconstruct the contents of input $x$ from $\mathit{D}_{\mathit{sim}}$, while the $\Lagr_{\mathit{VAE}}$ term tries to adjust the SAE output, $F_S(x)$, towards the distribution of $\mathit{D}_{\mathit{real}}$ that was previously used for training the VAE.  This way, the SAE model learns to transform images from $\mathit{D}_{\mathit{sim}}$ into images similar to images from $\mathit{D}_{\mathit{real}}$.  To translate images from $\mathit{D}_{\mathit{sim}}$'s distribution using SAEVAE, we only use the SAE model and do not need to use the VAE model.

\textbf{Comparing translators:} SAEVAE has significantly less execution time, i.e., translation time, compared to the style transfer method. This is because, to infer images,  SAEVAE only  performs a forward pass of its SAE model, hence a short execution time. Further, SAEVAE has a lower training cost  compared to CycleGAN. This is because in SAEVAE training, VAE is frozen after the initial pretraining. That is, in Equation~\ref{fm:lvae}, only $F_S$ is trained and $F^*_V$ does not change. Hence, in contrast to CycleGAN, we do not train two complex DLs with competing goals simultaneously. Specifically, in our experiments, we observed that the average execution time 
for SAEVAE, CycleGAN, and neural style transfer are, respectively, $0.05$~sec, $0.30$~sec, and $4.69$~sec. That is, SAEVAE's execution time is noticeably better than that of both CycleGAN and neural style transfer. Further, as we show in Section~\ref{sec:res} and in \textbf{RQ2} and \textbf{RQ3}, SAEVAE is more effective than CycleGAN and neural style transfer in improving ADS-DNNs' offline and online testing results when the ADS-DNNs have test and training datasets from different domains.

\section{Online Testing with Translators}
\label{sec:approach}

\begin{figure*}[t]
    \centering
    \includegraphics[width=0.9\textwidth]{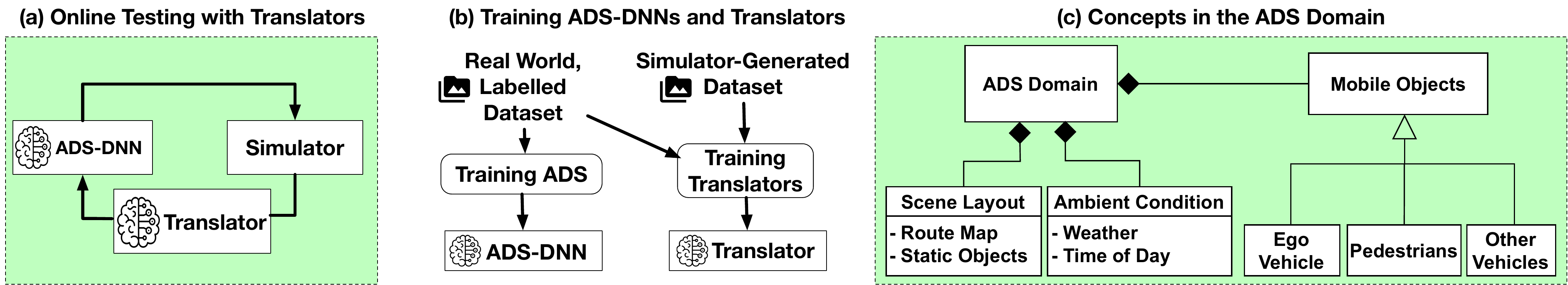} 
    \vspace*{-.2cm}
    \caption{Extending online testing for ADS with translators: (a)~The embedding of translators into the online testing loop;  (b)~prerequisite datasets for training ADS-DNNs and translators; and (c)~overview of concepts in the ADS domain} 
    \label{fig:sut}
    \vspace*{-.2cm}
\end{figure*}

Figure~\ref{fig:sut}(a) shows the augmentation of online ADS testing using translators. A translator is integrated within the online testing loop, which comprises both a simulator and an ADS-DNN. Every image produced by the simulator is fed to this translator, which then transforms it to align with the distribution of the training set of the ADS-DNN. The images generated by the translator are passed to the ADS-DNN to  determine its output, e.g., a steering angle value.  For example, Figure~\ref{fig:samples}(c)  shows  the transformation of the simulator-generated image in Figure~\ref{fig:samples}(b) by SAEVAE.  As the figures show, the transformed image remains similar to the original synthetic image. However, 
although it might not be obvious to our eyes, the transformed image aligns better with the real-world image than the original synthetic image. We demonstrate this alignment through a metric that measures the closeness of image distributions in RQ1.

Figure~\ref{fig:sut}(b) illustrates the prerequisites for the translator-based online testing   shown in  Figure~\ref{fig:sut}(a). These prerequisites include two datasets: a real-world dataset, which is the training set of the ADS-DNN under test, and a simulator-generated dataset for testing. The real-world dataset is required to be labelled, while the simulator-generated dataset is not.  Both datasets are required for training translators where the simulator-generated dataset is the source, and the real-world dataset is the target.  The images in the real-world and the simulator-generated datasets do not need to be paired, but they are expected to have been developed to assess the same ADS task and should include the same concepts from the ADS domain.  Figure~\ref{fig:sut}(c) illustrates key concepts in the ADS domain, including various objects, ambient conditions, and route layouts. ADS simulators can be configured to include or exclude these elements or modify their appearances, enabling the generation of synthetic images that resemble those in ADS-DNN training datasets.

\section{Evaluation Setup}
\label{sec:evaluation}
Our study is motivated by the increasing need to test ADS-DNNs, trained with real-world images, on synthetic images. When the training and test datasets have different distributions, translators are proposed to bring the datasets' distributions closer. 
In this section, we assess the effectiveness, reliability and efficiency of translators for ADS
testing.  Our experiments begin with \textbf{RQ1}, which assesses the effectiveness of translators in reducing the distribution gap between the real-world and synthetic datasets. \textbf{RQ2} and \textbf{RQ3}, respectively, examine the effectiveness of translators in improving the offline and online test accuracy results. To ensure the reliability of translators for ADS testing, \textbf{RQ4} evaluates whether translators can preserve test data quality  in terms of coverage, diversity, and fault-revealing ability. To assess the efficiency of translators, \textbf{RQ5} evaluates their time overhead in online testing.  Finally, \textbf{RQ6}  examines whether translators  can help  reduce the cost of ADS testing.

\textbf{RQ1. (Training and test data distributions gap mitigation)} \emph{How effectively do translators mitigate the distribution gap between training and test data for ADS-DNNs?} To properly evaluate ADS-DNNs, it is crucial to use test data that shares a similar distribution with their training data. This research question investigates how well translators align the distribution of synthetic test data with the distribution of the real-world data used for training them.

\emph{Metrics:}  To compare the distribution of a given real-world dataset with its corresponding synthetic dataset, both before and after applying translators, we train a VAE~\cite{vae} on the real-world dataset. We then compute the VAE reconstruction error for each image in the real-world dataset and in the synthetic dataset, both before and after translation. We use the mean absolute error (MAE) to calculate the reconstruction errors. A VAE trained on a real-world dataset likely  exhibits larger reconstruction errors for non-real-world images~\cite{vae, daiv, self_oracle}. We leverage this characteristic to compare the similarity between real-world and synthetic datasets.

\textbf{RQ2. (Offline testing accuracy gap mitigation)} \emph{How effectively do translators mitigate the accuracy gap  in offline ADS testing when training and test datasets are from different domains?} This research question explores whether there is a decline in accuracy of ADS-DNNs when they are tested on synthetic images instead of real-world images. We then examine how well translators
reduce this accuracy decline.

\emph{Metrics:} We evaluate the accuracy of ADS-DNNs for real-world and synthetic datasets. For the lane-keeping task, we use MAE, and for the object-detection task, we use the mean Average Precision (mAP), a widely recognized metric for object-detection ~\cite{map, coco}.

\textbf{RQ3. (Online testing failure reduction)} \emph{How effectively do translators reduce the occurrence and severity of failures in online ADS testing?} ADS-DNNs may exhibit failures during online testing due to the mismatch between the distributions of the simulator-generated images and their real-world training-set images. We investigate if the translators can reduce the occurrence or the severity of such failures in online testing.

\emph{Metrics:} We use both the out-of-bound (OOB) distance and the number of out-of-bounds (OOBs) to assess online testing results~\cite{sbftgithub, two_is_better,Biagiola_sbft, Arcaini_crag, Riccio_deepjanus}. The OOB distance measures the average difference between the lane's width and the distance from the car to the lane's center across all time steps of each scenario. The OOB distance is maximized when the ego vehicle is centered in the lane and decreases with any deviation. A higher OOB distance value indicates better performance of the ADS-DNN in maintaining the vehicle closer to the lane center on average. In addition to the OOB distance, we report the number of out-of-bounds. An out-of-bounds occurs whenever the percentage of the ego vehicle that is outside the lane exceeds a specific threshold. We use a threshold of 0.5, meaning an OOB occurs when at least half of the car moves outside the lane. In the remainder of this paper, we refer to the \emph{OOB distance} as \emph{OOB} and to \emph{the number of OOBs} as \emph{\#OOB}, respectively.

\textbf{RQ4. (Test data quality preservation)} \emph{Can translators preserve the test data quality in terms of diversity, coverage and fault-revealing ability?} We investigate whether using translated data for testing may result in the loss of any qualities of the original test data in exercising ADS-DNNs.

\emph{Metrics:} For test-data diversity, we use the geometric diversity (GD)~\cite{geometric} and standard deviation (SD)~\cite{clustering}. For test coverage, we use neuron coverage (NC)~\cite{neuron_coverage} and  surprise adequacy (SA)~\cite{surprise}. For the fault-revealing ability of test data, we use a commonly-used latent space clustering approach~\cite{clustering}.

\textbf{RQ5. (Translators' time overhead)} \emph{Is the time overhead incurred by using translators during online testing justified?}   A translator is time-efficient  for online testing if its execution time  is substantially smaller than the simulation time.

\emph{Metrics:} We evaluate the time required, during online testing, to process each simulation frame with and without translators.

\textbf{RQ6. (Online versus offline results correlation)} \emph{Can translators  reduce the cost of online testing by increasing the correlation between offline and online testing results?}  We investigate the effect of using translators in offline and online testing on the correlation between their results. If the correlation increases, the results of online testing --  the more expensive approach --  can be more accurately predicted from offline testing --  the less expensive approach -- hence, reducing the overall cost of ADS testing.

\emph{Metrics:} To calculate the correlation between offline and online testing results, we use Pearson and Spearman correlations \cite{correlations}. As discussed in RQ2, offline testing results are measured using MAE, and as discussed in RQ3, online testing is measured using the OOB and \#OOB metrics.

\vspace*{-0.2cm} 

\subsection{Offline Datasets and ADS-DNNs} 
\label{subsec:setupdnn}
Below, we first introduce the offline datasets for the lane-keeping and object-detection ADS tasks. For each task, we use a training dataset representing real-world features and a synthetic, test dataset generated by a simulator. We then describe different ADS-DNNs used to automate the tasks of lane keeping and object detection.

\textbf{Lane-keeping datasets.} For the lane-keeping training dataset, we use the Udacity Jungle dataset~\cite{udacity_jungle}. 
While this dataset is not taken from the real world, it is commonly used to train ADS-DNNs for the lane-keeping task because its images closely resemble real-world images. Furthermore, as we show in RQ1, the image distribution of the Udacity Jungle dataset is distinct from those generated by BeamNG, our ADS simulator, making this dataset a suitable match for our experiments. Each Udacity Jungle image  is labelled with a ground-truth steering angle obtained from a human driver.

We generate the synthetic lane-keeping dataset corresponding to the  Udacity Jungle dataset, using  the BeamNG simulator~\cite{beamng}.  We configure the BeamNG simulator to ensure that both the road and landscape match with those in the images of the Udacity Jungle dataset.  To compute MAE results for RQ2 and RQ6, we need ground-truth steering angle labels for a test-set split of synthetic images. For RQ2, we automatically label the synthetic dataset generated by BeamNG with steering angles obtained from BeamNG's AI controller (BeamNG.AI). Since, for the  RQ6 experiments, the ego vehicle is controlled by an ADS-DNN instead of BeamNG.AI, BeamNG does not provide ground-truth steering angles. Hence, we use an independent PID controller to label images for RQ6. For further details, see Section~\ref{subsec:intval}. The synthetic lane-keeping dataset is provided in our  in our replication package~\cite{github}.

\textbf{Object-detection datasets.} We use KITTI~\cite{kitti} and virtual KITTI (vKITTI)~\cite{vkitti}, respectively, for the real-world and synthetic object-detection datasets. The KITTI benchmark is extensively used for object detection, while vKITTI, its synthetic counterpart, was developed using the Unity game engine~\cite{unity}. The images in KITTI and vKITTI show a moving vehicle in various road environments.

\textbf{Lane-keeping ADS-DNNs.} We consider four different ADS-DNNs: Dave2, Autumn, Chauffeur and Epoch~\cite{dave2, udacity:challenge} for the lane-keeping task. These four ADS-DNNs are end-to-end regression models and produce steering angles to maneuver an ego vehicle. 
  
\textbf{Object-detection ADS-DNNs.} We use "You Only Look Once" (YOLOv5) \cite{yolov5}, a widely used DNN for object detection and classification. According to YOLO's GitHub, version 5 is the most stable and widely adopted version of YOLO.

\subsection{Training ADS-DNNs and Translators} 
\label{translator-setup}
As per the process shown in Figure~\ref{fig:sut}(b), we train the ADS-DNNs and the translators in our study using our real-world datasets for lane-keeping and object-detection tasks. 
Table~\ref{tab:train} presents the properties of the datasets and how we split them for training ADS-DNNs and translators, and for answering research questions. Specifically, we split the real-world datasets into $70$\% for training and tuning ADS-DNNs, and $30$\% for testing, i.e., evaluating ADS-DNNs' accuracy on the real-world dataset.
We train the ADS-DNNs with  $50$ epochs, batch size of 128 as per the standard practices~\cite{dlbook}.

\begin{table}[t]
\centering
\caption{Properties of the datasets used for training ADS-DNNs and translators, and for answering research questions}
\label{tab:train}
\scalebox{0.7}{
\begin{tabular}{|ll|ll|ll|}
\hline
\multicolumn{2}{|l|}{\multirow{2}{*}{\textbf{Task}}}                                                               & \multicolumn{2}{c|}{\textbf{Training and Validation Set}}                                                                                                    & \multicolumn{2}{c|}{\textbf{Test Set}}                                                                                                            \\ \cline{3-6} 
\multicolumn{2}{|l|}{}                                                                                    & \multicolumn{1}{l|}{\begin{tabular}[c]{@{}l@{}}\textbf{ADS-DNN}\\ \textbf{(labelled)}\end{tabular}} & \begin{tabular}[c]{@{}l@{}}\textbf{Translator}\\ \textbf{(unlabelled)}\end{tabular} & \multicolumn{1}{l|}{\begin{tabular}[c]{@{}l@{}}\textbf{RQ2, RQ6}\\ \textbf{(labelled)}\end{tabular}} & \begin{tabular}[c]{@{}l@{}}\textbf{RQ1, RQ4, RQ5}\\ \textbf{(unlabelled)}\end{tabular} \\ \hline
\multicolumn{1}{|l|}{\multirow{2}{*}{\begin{tabular}[c]{@{}l@{}}\textbf{Lane} \\ \textbf{Keeping}\end{tabular}}}     & \textbf{real} & \multicolumn{1}{l|}{\cellcolor{cyan!20}2363}                                                        &  3376                                                             & \multicolumn{1}{l|}{\cellcolor{cyan!20}1013}                                                    & 3376                                                      \\ \cline{2-6} 
\multicolumn{1}{|l|}{}                                                                             & \textbf{sim}  & \multicolumn{1}{l|}{N/A}                                                         &  8563                                                             & \multicolumn{1}{l|}{\cellcolor{green!20}8563}                                                    & 8563                                                      \\ \hline
\multicolumn{1}{|l|}{\multirow{2}{*}{\begin{tabular}[c]{@{}l@{}}\textbf{Object} \\ \textbf{Detection}\end{tabular}}} & \textbf{real} & \multicolumn{1}{l|}{\cellcolor{cyan!20}6358}                                                        & 7481                                                             & \multicolumn{1}{l|}{\cellcolor{cyan!20}1123}                                                    & 7481                                                      \\ \cline{2-6} 
\multicolumn{1}{|l|}{}                                                                             & \textbf{sim}  & \multicolumn{1}{l|}{N/A}                                                         & 2066                                                             & \multicolumn{1}{l|}{\cellcolor{cyan!20}2066}                                                    & 2066                                                      \\ \hline

\end{tabular}

}
\footnotesize{
\textbf{Note regarding the origin of the labels for the datasets:} The labelled datasets highlighted by \colorbox{cyan!20}{\phantom{00}}  are publicly available sets with pre-labeled images. The labelled dataset highlighted by \colorbox{green!20}{\phantom{00}} includes images generated by the BeamNG simulator and automatically labelled by controllers described in detail in  Section~\ref{subsec:intval}.}
\end{table}

As for the translators, we train SAEVAE, CycleGAN and neural style translators for both lane-keeping and object-detection tasks to bring the synthetic images closer to the distribution of real-world images. We train six translator models developed based on three different translator approaches discussed in Section \ref{sec:saevae} and for two different ADS tasks.  Note that the datasets used for training translators do not need to be labelled.

We tune ADS-DNNs and translators based on a $30\%$ random split from their training set as a validation set. For ADS-DNNs, we use a grid search for tuning and train several models and keep the best one yielding better loss on the validation set. For translators, we tune hyperparameters using grid search  and initialize the weights using Xavier initialization~\cite{Glorot_Xavier} to reduce randomness and increase robustness. We then train five SAEVAEs and five CycleGANs. We stage their training by saving model weights at each epoch and record the model with the lowest validation loss. For SAEVAE, the loss for the trained models show negligible differences. This is due to the more stable training of SAEVAE, since unlike CycleGAN, the training of SAE and VAE is not intertwined. Training SAEVAE is similar to training any simple DNN, e.g., CNNs. For CycleGAN, we select the model with the lowest validation loss. For neural style translators, we choose  several random images from the real-world datasets as style images. Among the resulting neural style translator models, we keep the one yielding the lowest MAE and the highest mAP values on the synthetic test set.

\subsection{Setup for Online ADS Testing}
\label{subsec:setuponline}
We use the BeamNG simulator~\cite{beamng} for online testing of ADS-DNNs and to answer RQ3, RQ5, and RQ6. Specifically, we use the ADS testing framework provided by the \emph{Cyber-Physical Systems Testing Tool Competition} track of the SBFT workshop~\cite{sbftgithub}. This framework monitors the vehicle’s ability to navigate the road while keeping within its lane. This framework provides alternative test input and road generators to challenge ADS-DNN’s maneuvering~\cite{Riccio_deepjanus, Arcaini_crag, naiverandom}. Our goal, however, is
to assess translators’ performance on a diverse set of roads instead
of exercising the maneuvering capabilities of ADS-DNNs. Hence,
we use, among these road generators, the naive random road generator~\cite{naiverandom}, which generates roads by interpolating four randomly
generated waypoints. We discard all invalid roads and ensure that
all the roads used in our study are valid. We set the ego vehicle's speed to $30$km/h to be consistent with the average speed of the ego vehicle in the Udacity Jungle dataset. In addition, the generated roads are two-lane rural roads passing through a green landscape without any other vehicles, buildings, pedestrians, or objects. This ensures that the scenes produced by BeamNG are similar to the images in the Udacity Jungle dataset. The generated roads are
available as part of our replication package~\cite{github}. We use the generated roads to test  our four end-to-end ADS-DNNs, i.e., Autumn, Chauffeur, Dave2 and Epoch with and without a translator.  Since YOLO is not an end-to-end ADS, it cannot drive a vehicle on its own. Hence, it cannot be used within this online test setup.

\subsection{Test Data Quality Metrics}
\label{sec:metrics}
We consider test data quality metrics to measure coverage, diversity, and fault-revealing capability. These metrics can be white box or black box. White-box metrics use the architecture of the DNN under test, while black-box metrics are independent of the DNN's architecture. In our evaluation, we compute the black-box metrics for our synthetic datasets before and after translation, and the white-box metrics for each pair of an ADS-DNN and a synthetic dataset before and after translation. The metrics are described below. 

\textbf{White-Box Metrics.} We assess the coverage of each test set for each  DNN under test using neuron coverage (NC)~\cite{neuron_coverage} and surprise adequacy (SA)~\cite{surprise}. Given a test set and a DNN, NC measures the percentage of neurons of the DNN whose activation levels exceed a specified threshold. We chose $0$ for the threshold as suggested in the original paper introducing the NC metric~\cite{neuron_coverage}. 
To compute SA, given a test set and a DNN, we measure the percentage of the test inputs in the test set that are surprising for the DNN. To determine if a test input is surprising for a DNN, we calculate its surprise amount, defined as the Mahalanobis distance between the test input's latent space vector and the distribution of latent space vectors from the DNN's training set.  A test input is surprising if its corresponding Mahalanobis distance is larger than a specified threshold. We choose the threshold to be one standard deviation above the mean of the surprise distribution of the DNN's training set to ensure that the majority of the training set images are not considered surprising for the DNN. Note that our preliminary results show that different thresholds for the SA metric do not change the relative comparison of test datasets' coverage  before and after translation. More precisely, the choice of the SA threshold does not affect the relative differences in SA values, provided that the majority of the training set is considered unsurprising and consistent thresholds are used for both sides of the comparison.

To assess the fault-revealing ability of a given test set for a DNN under test, we compute the number of clusters in the latent space vectors obtained by feeding the test inputs in the test set to the DNN. 
This measure for the fault-revealing ability of the DNN's test data is adopted from previous work~\cite{clustering, clustering:GIST}. The intuition behind this measure is that misbehavior-inducing test inputs that form a cluster in the latent space of the DNN under test likely point to the same fault. Hence, the number of clusters obtained for each test set can serve as a proxy for the number of faults that the test set can reveal.

\textbf{Black-Box Metrics.} We use geometric diversity (GD)~\cite{geometric} and standard deviation (SD)~\cite{clustering} to measure test data diversity. Given a test dataset, GD measures the hypervolume spanned by the latent space vectors of the dataset. These latent space vectors are derived by feeding the test set to a pretrained DNN, different from the DNN under test. We use VGG19~\cite{vgg19}, a well-known vision model, for this purpose. The main idea behind GD is that the latent space vectors are compact representations of the high-level features of the test data, and can be used to assess how diversely the test set spans the feature space and  exercises the DNN under test.
For SD, we construct a vector in which each component represents the standard deviation of that component across all latent space vectors generated during GD calculations. The SD metric is the Euclidean norm of this vector. The goal of SD is to quantify the variability within the feature space of the test data.

\section{Results}
\label{sec:res}
In this section, $D_{\mathit{real}}$ denotes our real-world datasets, and 
$D_{\mathit{sim}}$ denotes our synthetic datasets. As shown in Table~\ref{tab:train}, $D_{\mathit{sim}}$ is exclusively used for testing, whereas $D_{\mathit{real}}$ is partitioned for both training and testing. For our results, we specify whether the training or testing partitions of $D_{\mathit{real}}$ are used, denoting them by $D_{\mathit{real, train}}$ and $D_{\mathit{real, test}}$, respectively.  Further, the CycleGAN and neural style translators are referred to as cycleG and styleT, respectively.

\subsection{RQ1: Test data distribution gap mitigation} 
We investigate if translators can effectively mitigate the distribution gap between  the $D_{\mathit{real}}$ test data and the $D_{\mathit{sim}}$ test data. For each task, we train a VAE on the $D_{\mathit{real}}$ training data. We tune the VAE according to the tuning process described in Section~\ref{translator-setup} for our ADS-DNNs and translators. We then use the trained VAE to reconstruct $D_{\mathit{real}}$ and $D_{\mathit{sim}}$ test sets as well as the translations of  $D_{\mathit{sim}}$ obtained by the SAEVAE, cycleG and styleT translators. 

\begin{figure}[t]
     \centering
     \begin{subfigure}{0.99\columnwidth}
         \centering
     \includegraphics[width=0.9\columnwidth]{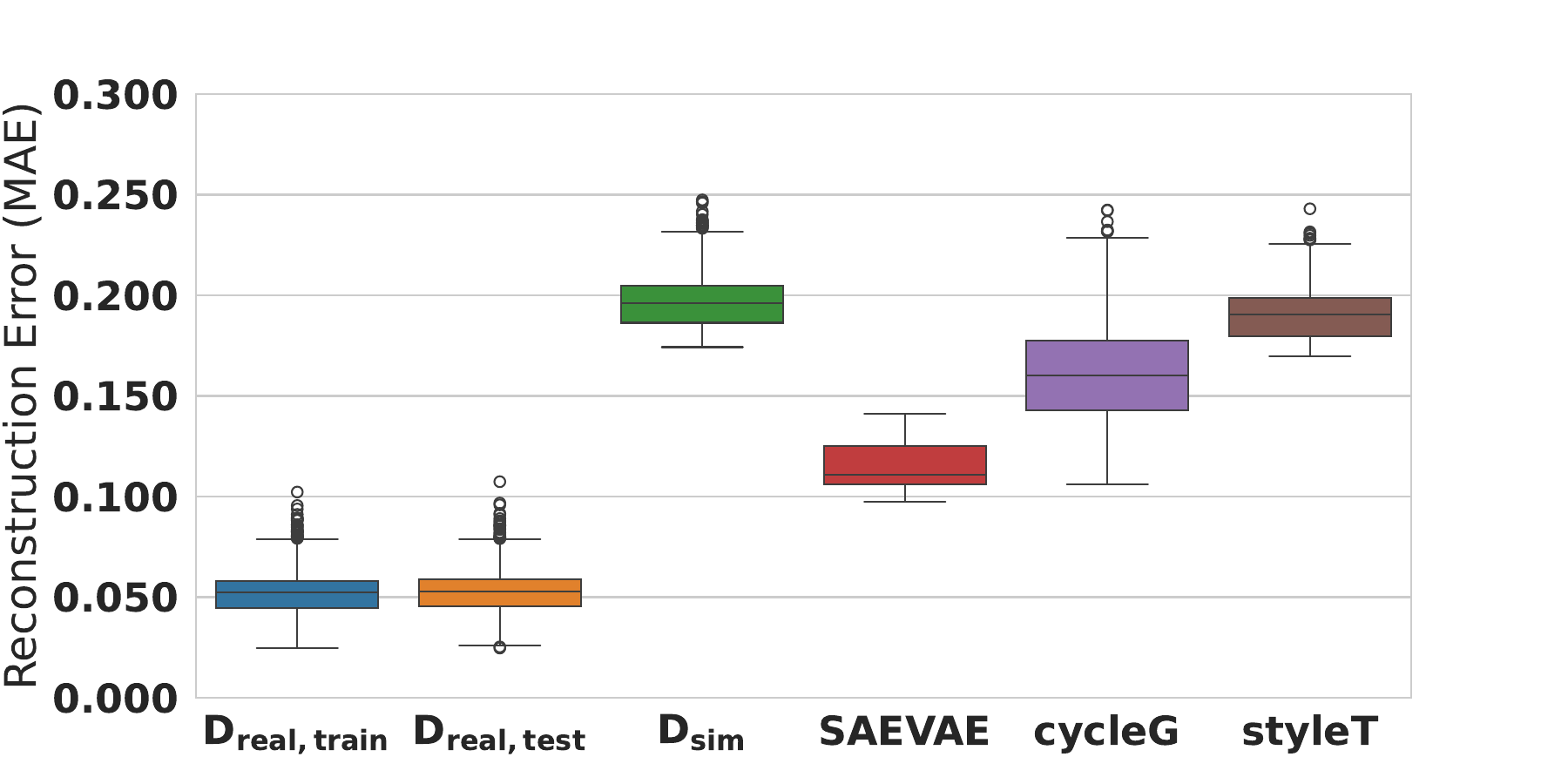}
     \caption{Lane-Keeping Task}
     \label{fig:rq2_reconstruction_box_lane}
     \end{subfigure}
     \hfill
     \begin{subfigure}{0.99\columnwidth}
         \centering
     \includegraphics[width=0.9\columnwidth]{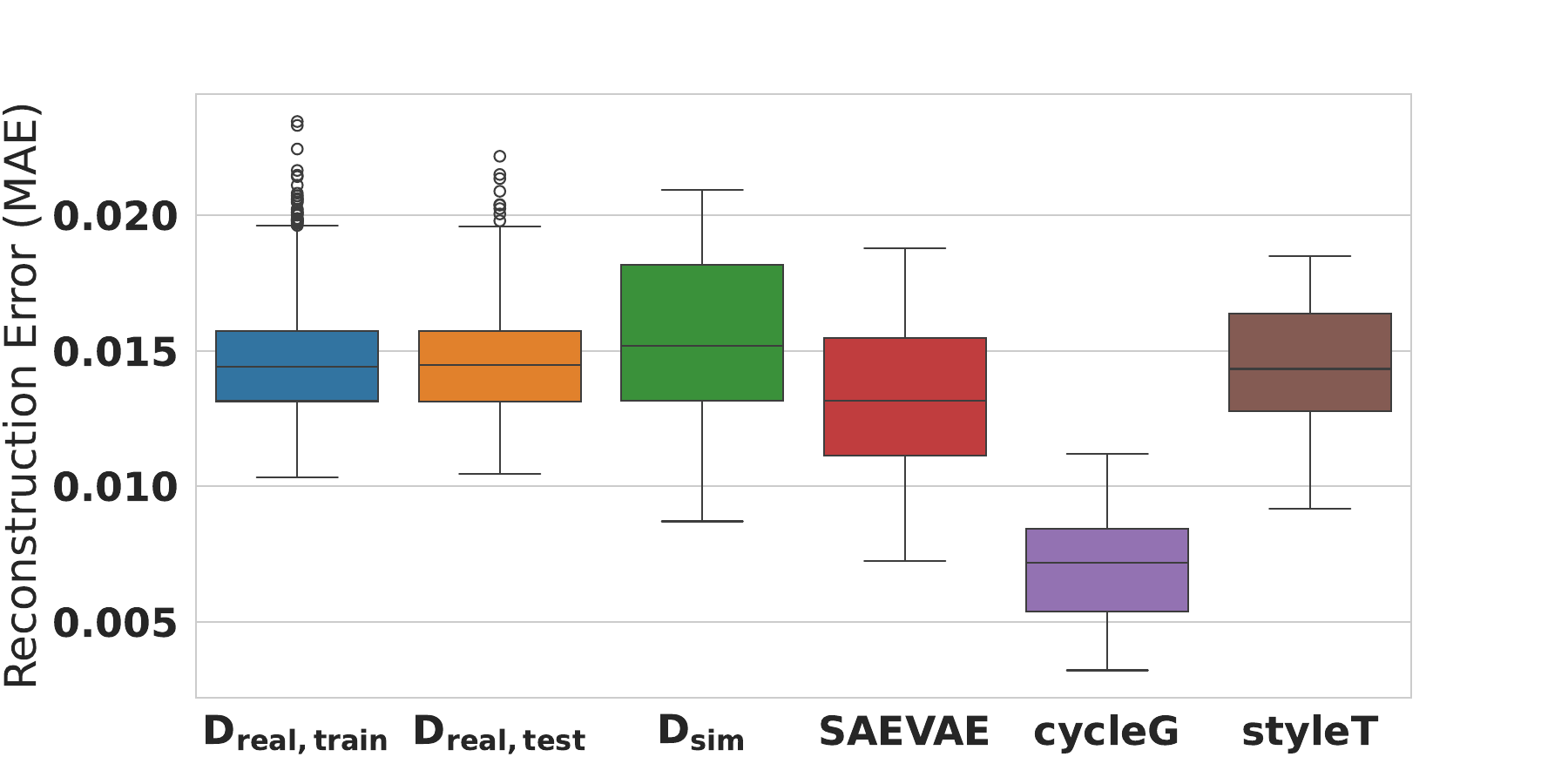}
     \caption{Object-Detection Task}
     \label{fig:rq2_reconstruction_box_obj}
     \end{subfigure}
     \hfill
        \caption{Reconstruction-error distributions obtained by a VAE trained on $D_{\mathit{real}}$ for lane-keeping and object-detection tasks. The error distributions are shown for $D_{\mathit{real}}$, $D_{\mathit{sim}}$, and the translations of $D_{\mathit{sim}}$ by SAEVAE, cycleG and styleT.}
    \label{fig:rq2_reconstruction_box}
        
\end{figure}

Figure~\ref{fig:rq2_reconstruction_box} compares the reconstruction-error distributions for the six different datasets for both lane-keeping and object-detection tasks. 
As expected, $D_{\mathit{real}}$ training and test sets have very close reconstruction-error distributions.  
Figure~\ref{fig:rq2_reconstruction_box}(a) shows a significant difference between the distributions of the $D_{\mathit{sim}}$ and $D_{\mathit{real}}$ test sets for the lane-keeping task. Further, it shows that SAEVAE can reduce the gap between the $D_{\mathit{sim}}$ and $D_{\mathit{real}}$ distributions more effectively compared to cycleG and styleT. As for the object-detection task, since the differences between the box plots in Figure~\ref{fig:rq2_reconstruction_box}(b) are not visually distinct, we report the statistical-test results in Table~\ref{tab:rq2_stattest}. These results compare the data distribution of the $D_{\mathit{real}}$ test set with the data distributions of $D_{\mathit{sim}}$ and those obtained by SAEVAE, cycleG, and styleT. The statistical tests confirm that SAEVAE produces a dataset with reconstruction errors not significantly different from those of $D_{\mathit{real}}$. This shows the similarity between the SAEVAE-translated synthetic dataset and the $D_{\mathit{real}}$ test dataset.

\begin{table}[t]
\centering
\vspace*{.3cm}
\caption{Statistical tests (Wilcoxon and Vargha-Delaney) comparing the VAE reconstruction errors between  $D_{\mathit{real,test}}$ and other datasets underlying Figure~\ref{fig:rq2_reconstruction_box}(b)}
\label{tab:rq2_stattest}
\vspace*{-.25cm}
\scalebox{0.85}{
\begin{tabular}{|l|l|l|l|l|}
\hline
                 & \textbf{$\textbf{D}_{\textbf{sim}}$} & \textbf{SAEVAE} & \textbf{cycleG} & \textbf{styleT} \\ \hline
\textbf{Effect size ($A_{12}$)}     & 0.78 (L)          & 0.51          & 1 (L)               & 0.67 (M)          \\ \hline
\textbf{p-value} & $\sim 0$       & 0.98            & $\sim 0$       & $\sim 0$       \\ \hline
\end{tabular}}
\end{table}

\begin{tcolorbox}[breakable, colback=gray!10!white,colframe=black!75!black]
\textbf{RQ1:} For the lane-keeping task, all the translators reduce the gap between the real-world and synthetic test sets. For the object-detection task, only SAEVAE and styleT decrease this gap,  with SAEVAE outperforming cycleG and styleT in reducing the distribution gaps for both tasks.
\end{tcolorbox}

\subsection{RQ2: Offline testing accuracy gap mitigation}
To evaluate the effectiveness of  translators, i.e., SAEVAE, cycleG and styleT, in reducing the offline testing accuracy gap, we first transform each $D_{\mathit{sim}}$ set for  each ADS task  using each translator. This process yields  six different translated test sets, i.e., three  for each ADS task.  We then apply our five ADS-DNNs -- four for lane keeping  and one for object detection -- to their respective $D_{\mathit{real,test}}$, $D_{\mathit{sim}}$, and the six translated test sets.

\begin{figure}[t]
    \centering
    \includegraphics[width=1\columnwidth]{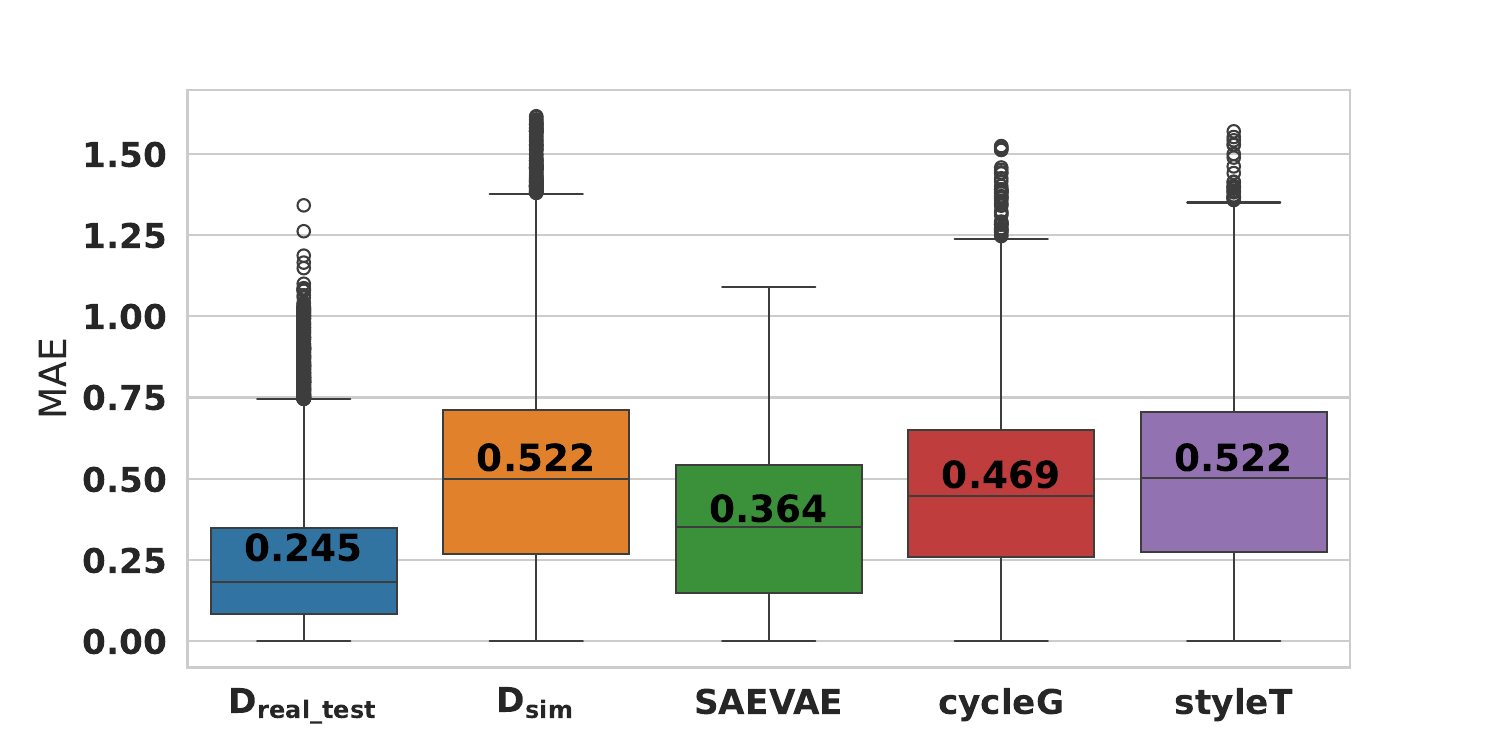}
    \caption{MAE results of the lane-keeping ADS-DNNs for the real-world ($D_{\mathit{real, test}}$) and  synthetic  ($D_{\mathit{sim}}$) datasets, and for the translations of  $D_{\mathit{sim}}$ obtained by SAEVAE, cycleG and styleT}
    \label{fig:rq1_offline_boxplot}
\end{figure}

Figure~\ref{fig:rq1_offline_boxplot} shows the MAE results assessing the accuracy of  the lane-keeping task of our four ADS-DNNs applied to $D_{\mathit{real,test}}$, $D_{\mathit{sim}}$, and the three translations of $D_{\mathit{sim}}$ by SAEVAE, cycleG and styleT, respectively.  A lower MAE indicates better lane-keeping accuracy. As expected, the accuracy of $D_{\mathit{sim}}$ is significantly degraded compared to that of $D_{\mathit{real,test}}$, indicating a substantial decline in accuracy for synthetic images compared to real-world images. Among the three translations of $D_{\mathit{sim}}$, the one obtained by SAEVAE  has the best accuracy, i.e., lowest MAE. Table~\ref{tab:rq1_stattest} compares the MAE results in Figure~\ref{fig:rq1_offline_boxplot} using the Wilcoxon statistical test and the Vargha-Delaney effect size. The table shows that the MAE results obtained by applying SAEVAE to $D_{\mathit{sim}}$ are significantly better than those for $D_{\mathit{sim}}$ before translation or with other translations. 

\begin{table}[t]
\centering
\vspace*{.3cm}
\caption{Statistical-test results comparing the MAE distributions in Figure~\ref{fig:rq1_offline_boxplot}}
\label{tab:rq1_stattest}
\vspace*{-.25cm}
\scalebox{0.72}{
\begin{tabular}{|l|p{2.2cm}|p{2.8cm}|p{2.8cm}|}
\hline
 & \textbf{SAEVAE vs No Translator} & \textbf{SAEVAE vs cycleG} & \textbf{SAEVAE vs styleT} \\ \hline
\textbf{p-value}          & $\sim 0$              & $\sim 0$         & $\sim 0$         \\ \hline
\textbf{Effect size ($A_{12}$)}   & $0.35$ (M)                    & $0.38$ (M)             & $0.33$ (M)             \\ \hline
\end{tabular}}
\end{table}

Table~\ref{tab:rq1_obj_offline} shows the mAP results evaluating the accuracy of  the object-detection task for $D_{\mathit{real,test}}$, $D_{\mathit{sim}}$, and the three datasets obtained by SAEVAE, cycleG and styleT, respectively. The mAP results are obtained by assessing if our object-detection ADS-DNN can identify approximately 24K car instances in the  datasets. 
Note that we only assess the object-detection task for the car class, since other classes common to both KITTI and vKITTI have significantly fewer instances (i.e., less than 2,000), which is insufficient for training and assessing a model.
Table~\ref{tab:rq1_obj_offline} reports both mAP50 and mAP50-95: mAP50 calculates the mAP metric at a 50\% IOU threshold, while mAP50-95 averages the mAP metric across IOU thresholds from 50\% to 95\%.  A higher mAP indicates better object-detection accuracy. Similar to the results for lane keeping, the object-detection results confirms that (1)~there is a significant accuracy gap between the real-world and synthetic images, and (2)~SAEVAE outperforms the other translators in reducing the accuracy gap between real-world and synthetic images. 

\begin{table}[t]
\centering
\vspace*{.3cm}
\caption{The accuracy of the object detection ADS-DNN for the real-world ($D_{\mathit{real}}$) and  synthetic  ($D_{\mathit{sim}}$) datasets, and for the translations of  $D_{\mathit{sim}}$ by SAEVAE, cycleG and styleT}
\label{tab:rq1_obj_offline}
\vspace*{-.25cm}
\scalebox{0.72}{
\begin{tabular}{|c|c|c|ccccc|}
\hline
\multirow{2}{*}{\textbf{Class}} & \multirow{2}{*}{\textbf{Instances}} & \multirow{2}{*}{\textbf{Metric}} & \multicolumn{5}{c|}{\textbf{Dataset}}                                                                                                             \\ \cline{4-8} 
                       &                            &                         & \multicolumn{1}{c|}{$\textbf{D}_{\mathit{\textbf{real, test}}}$}    & \multicolumn{1}{c|}{$\textbf{D}_{\mathit{\textbf{sim}}}$}   & \multicolumn{1}{c|}{\textbf{SAEVAE}}   & \multicolumn{1}{c|}{\textbf{cycleG}}   & \textbf{styleT}   \\ \hline
\multirow{2}{*}{Car}   & \multirow{2}{*}{24540}     & mAP50                   & \multicolumn{1}{c|}{0.0143}   & \multicolumn{1}{c|}{0.00651}  & \multicolumn{1}{c|}{0.00793}  & \multicolumn{1}{c|}{0.00376}  & 0.00538  \\ \cline{3-8} 
                       &                            & mAP50-95                & \multicolumn{1}{c|}{0.00306}  & \multicolumn{1}{c|}{0.00163}  & \multicolumn{1}{c|}{0.00183}  & \multicolumn{1}{c|}{0.00111}  & 0.00123  \\ \hline
\end{tabular}
}
\vspace*{0.2cm}
\end{table}

\begin{tcolorbox}[breakable, colback=gray!10!white,colframe=black!75!black]
\textbf{RQ2:} 
Two translators, SAEVAE and cycleG, reduce the accuracy gap between simulator-generated and real-world datasets in offline testing for the lane-keeping task, with SAEVAE outperforming cycleG in terms of reduction. Furthermore, at least one translator, SAEVAE, reduces this accuracy gap for the object-detection task.
\end{tcolorbox}

\subsection{RQ3: Online testing failure reduction} 
\label{subsec:rq3}
In this research question, we investigate if the translators can reduce the occurrence  or the severity of failures during online testing.  We generate $100$ test scenarios randomly.  Each scenario is executed  four times: initially without a translator, and subsequently with each of the three translators separately.

\begin{table}[t]
\centering
\caption{Average OOB and \#OOB for  four 
lane-keeping ADS-DNNs with and without translators. The values highlighted in blue represent the best OOB and \#OOB for each ADS-DNN}
\label{tab:rq4_online}
\vspace*{-0.25cm}
\scalebox{0.75}{
\begin{tabular}{|l|ll|ll|ll|ll|}
\hline
\multirow{2}{*}{\textbf{ADS DNN}} & \multicolumn{2}{l|}{\textbf{w/o Translators}}       & \multicolumn{2}{l|}{\textbf{with SAEVAE}}         & \multicolumn{2}{l|}{\textbf{with cycleG}}         & \multicolumn{2}{l|}{\textbf{with styleT}}         \\ \cline{2-9} 
                                  & \multicolumn{1}{l|}{\textbf{OOB}} & \textbf{\textbf{$\#\text{OOB}$}} & \multicolumn{1}{l|}{\textbf{OOB}} & \textbf{$\#\text{OOB}$} & \multicolumn{1}{l|}{\textbf{OOB}} & \textbf{$\#\text{OOB}$} & \multicolumn{1}{l|}{\textbf{OOB}} & \textbf{$\#\text{OOB}$} \\ \hline
\textbf{Autumn}                   & \multicolumn{1}{l|}{1.18}      & 2.10 & \multicolumn{1}{l|}{\cellcolor{cyan!20}2.13}  & \cellcolor{cyan!20}1.59      & \multicolumn{1}{l|}{1.46}  & \cellcolor{cyan!20}1.59      & \multicolumn{1}{l|}{1.34}  & 2.26      \\ \hline
\textbf{Chauffeur}                & \multicolumn{1}{l|}{1.15}      & 2.47 & \multicolumn{1}{l|}{\cellcolor{cyan!20}1.92}  & \cellcolor{cyan!20}1.47      & \multicolumn{1}{l|}{1.13}  & 2.15      & \multicolumn{1}{l|}{1.32}  & 2.6      \\ \hline
\textbf{Dave2}                    & \multicolumn{1}{l|}{1.6}         & 1.74 & \multicolumn{1}{l|}{\cellcolor{cyan!20}1.62}  & \cellcolor{cyan!20}1.31      & \multicolumn{1}{l|}{\cellcolor{cyan!20}1.61}  & 1.76      & \multicolumn{1}{l|}{1.56}  &  2.11      \\ \hline
\textbf{Epoch}                    & \multicolumn{1}{l|}{1.16}      & 1.91 & \multicolumn{1}{l|}{\cellcolor{cyan!20}1.59}  & \cellcolor{cyan!20}1.37      & \multicolumn{1}{l|}{1.48}  & \cellcolor{cyan!20}1.36      & \multicolumn{1}{l|}{1.31}  & 1.5      \\ \hline
\end{tabular}
}
\vspace*{0.2cm}
\end{table}


\begin{table}[t]
\centering
\caption{Statistical-test results  comparing the OOB and \#OOB results achieved with SAEVAE against those obtained by other translators and without any translator. The p-values are indicated by p in the table, and the effect sizes are labelled as Large (L), Medium (M), Small (S) and Negligible (N). The p-values highlighted in blue represent cases where SAEVAE significantly outperforms the compared alternative. The non-highlighted p-values indicate that neither SAEVAE nor the compared alternative statistically outperforms the other.}
\label{tab:rq3_stattest}
\vspace*{-0.2cm}
\scalebox{0.58}{
\begin{tabular}{|l|cccc|cccc|cccc|}
\hline
\multirow{3}{*}{\textbf{ADS DNN}} & \multicolumn{4}{c|}{\textbf{SAEVAE vs No Translator}}                                                                            & \multicolumn{4}{c|}{\textbf{SAEVAE vs cycleG}}                                                                                   & \multicolumn{4}{c|}{\textbf{SAEVAE vs styleT}}                                                                                   \\ \cline{2-13} 
                                  & \multicolumn{2}{c|}{\textbf{OOB}}                                     & \multicolumn{2}{c|}{\textbf{\#OOB}}                & \multicolumn{2}{c|}{\textbf{OOB}}                                     & \multicolumn{2}{c|}{\textbf{\#OOB}}                & \multicolumn{2}{c|}{\textbf{OOB}}                                     & \multicolumn{2}{c|}{\textbf{\#OOB}}                \\ \cline{2-13} 
                                  & \multicolumn{1}{c|}{\textbf{A12}} & \multicolumn{1}{c|}{\textbf{p}} & \multicolumn{1}{c|}{\textbf{A12}} & \textbf{p}& \multicolumn{1}{c|}{\textbf{A12}} & \multicolumn{1}{c|}{\textbf{p}} & \multicolumn{1}{c|}{\textbf{A12}} & \textbf{p}& \multicolumn{1}{c|}{\textbf{A12}} & \multicolumn{1}{c|}{\textbf{p}} & \multicolumn{1}{c|}{\textbf{A12}} & \textbf{p}\\ \hline
\textbf{Autumn}                   & \multicolumn{1}{c|}{0.79(L)}         & \multicolumn{1}{c|}{\cellcolor{cyan!20}0}                & \multicolumn{1}{c|}{0.46(S)}         & \cellcolor{cyan!20}0& \multicolumn{1}{c|}{0.88(L)}         & \multicolumn{1}{c|}{\cellcolor{cyan!20}0}                & \multicolumn{1}{c|}{0.46}         & 0.9& \multicolumn{1}{c|}{0.82(L)}         & \multicolumn{1}{c|}{\cellcolor{cyan!20}0}                & \multicolumn{1}{c|}{0.42(S)}         & \cellcolor{cyan!20}0\\ \hline
\textbf{Chauffeur}                & \multicolumn{1}{c|}{0.78(L)}         & \multicolumn{1}{c|}{\cellcolor{cyan!20}0}                & \multicolumn{1}{c|}{0.43(S)}         & \cellcolor{cyan!20}0& \multicolumn{1}{c|}{0.88(L)}         & \multicolumn{1}{c|}{\cellcolor{cyan!20}0}                & \multicolumn{1}{c|}{0.41-S}         & \cellcolor{cyan!20}0& \multicolumn{1}{c|}{0.82(L)}         & \multicolumn{1}{c|}{\cellcolor{cyan!20}0}                & \multicolumn{1}{c|}{0.42(S)}         & \cellcolor{cyan!20}0\\ \hline
\textbf{Dave2}                    & \multicolumn{1}{c|}{0.50}         & \multicolumn{1}{c|}{0.5}                & \multicolumn{1}{c|}{0.45(S)}         & \cellcolor{cyan!20}0& \multicolumn{1}{c|}{0.5}         & \multicolumn{1}{c|}{0.11}             & \multicolumn{1}{c|}{0.31-M}         & \cellcolor{cyan!20}0& \multicolumn{1}{c|}{0.53(N)}         & \multicolumn{1}{c|}{\cellcolor{cyan!20}0.005}                & \multicolumn{1}{c|}{0.36(M)}         & \cellcolor{cyan!20}0\\ \hline
\textbf{Epoch}                    & \multicolumn{1}{c|}{0.69(M)}         & \multicolumn{1}{c|}{\cellcolor{cyan!20}0}             & \multicolumn{1}{c|}{0.51}         & 0.8& \multicolumn{1}{c|}{0.54(S)}         & \multicolumn{1}{c|}{\cellcolor{cyan!20}0.001}                & \multicolumn{1}{c|}{0.57}         & 0.1& \multicolumn{1}{c|}{0.66(M)}         & \multicolumn{1}{c|}{\cellcolor{cyan!20}0}                & \multicolumn{1}{c|}{0.53}         & 0.3\\ \hline
\end{tabular}
}
\vspace*{0.2cm}
\end{table}

Table~\ref{tab:rq4_online} shows the average OOB and \#OOB over 100 scenarios for each lane-keeping ADS-DNN both without any translator and with each of the translators applied individually. Values highlighted in blue are the best, or within $0.01$ of the best, OOB and \#OOB values for each ADS-DNN. For OOB, higher values are better, and  for \#OOB, lower values are better. For both OOB and \#OOB, SAEVAE yields the best or near-best averages  for all ADS-DNNs. Further, in three cases, cycleG yields  OOB and \#OOB  averages that are on par with those obtained by SAEVAE. However, the  OOB and \#OOB averages obtained by styleT and no translator are never on par with those obtained by SAEVAE or cycleG.

Table~\ref{tab:rq3_stattest} shows the statistical test and effect-size results comparing the OOB and \#OOB distributions. For all ADS-DNNs except Dave2, the SAEVAE's OOB results are significantly better than those obtained without translators or with cycleG and styleT, with small, medium, or large effect sizes. Further, for all ADS-DNNs except Epoch, the SAEVAE's \#OOB results are significantly better than those obtained without translators or with styleT with small or medium effect sizes. Finally, the \#OOB and OOB results of SAEVAE are never significantly worse than those obtained by styleT, cycleG or with no translator.

\begin{tcolorbox}[breakable, colback=gray!10!white,colframe=black!75!black]
\textbf{RQ3:} On average, the occurrence and severity of failures during online testing, as measured by OOB and \#OOB, are reduced with SAEVAE compared to using other translators or not using a translator. Further, statistical tests show that, compared to not using any translators, employing SAEVAE during online testing either significantly reduces or does not significantly increase the occurrence and severity of failures, as measured by OOB and \#OOB.
\end{tcolorbox}

\subsection{RQ4: Test data quality preservation} 
\label{subsec:rq4}
For this research question, we compare the test quality metrics described in Section~\ref{sec:metrics} for our synthetic test sets, $D_{\mathit{sim}}$, for both ADS tasks before and after translation.  
Table \ref{tab:rq4_ncsa} reports the differences between the  coverage that the $D_{\mathit{sim}}$ test sets achieve after translation by SAEVAE, cycleG, and styleT for our five ADS-DNNs compared to the coverage they achieve for the same ADS-DNNs before translation.  The coverage differences are in terms of NC and SA, defined in Section~\ref{sec:metrics}, and shown as percentages, calculated by subtracting the pre-translation coverage from the post-translation coverage since NC and SA values are normalized from zero to one. The positive percentage differences in Table \ref{tab:rq4_ncsa} indicate preserved or improved coverage and  are highlighted green. The negative percentage differences indicate decreased coverage and are  highlighted yellow. For all the ADS-DNNs, at least one translator preserves the coverage quality of the original test dataset. Further, except for three out of $30$ values in Table \ref{tab:rq4_ncsa}, the coverage quality after translation is either improved or preserved, or is decreased by less than $10$\%. As for SAEVAE, our best-performing translator, except for the SA metric for Epoch, the coverage quality is always preserved or decreases negligibly by less than 3\%.

\begin{table}[t]
\centering
\caption{The differences in test coverage obtained by the $\text{D}_{\text{sim}}$ test sets before and after translation in terms of Neuron Coverage (NC) and Surprise Adequacy (SA): $\approx$ means the test coverage is preserved by the translator; ↓ indicates that the translator decreases the test coverage; and  ↑  indicates that the translator improves the test coverage }
\label{tab:rq4_ncsa}
\vspace*{-0.25cm}
\scalebox{0.75}{
\begin{tabular}{|l|c|c|c|c|c|c|c|c|c|c|}
\hline
\multirow{2}{*}{Translator} & \multicolumn{2}{c|}{Autumn} & \multicolumn{2}{c|}{Chauffeur} & \multicolumn{2}{c|}{Dave2} & \multicolumn{2}{c|}{Epoch} & \multicolumn{2}{c|}{YOLO} \\ \cline{2-11} 
                         & NC & SA & NC & SA & NC & SA & NC & SA & NC & SA \\ \hline
SAEVAE                   & \cellcolor{green!25}$\approx$ & \cellcolor{yellow!25}↓ 3\% & \cellcolor{yellow!25}↓ 2\% & \cellcolor{green!25}$\approx$ & \cellcolor{green!25}$\approx$ & \cellcolor{green!25}$\approx$ & \cellcolor{yellow!25}↓ 3\% & \cellcolor{yellow!25}↓ 39\% & \cellcolor{green!25}↑ 2\% & \cellcolor{yellow!25}↓ 3\% \\ \hline
cycleG                   & \cellcolor{green!25}$\approx$ & \cellcolor{green!25}↑ 1\% & \cellcolor{green!25}↑ 5\% & \cellcolor{green!25}$\approx$ & \cellcolor{green!25}$\approx$ & \cellcolor{yellow!25}↓ 96\% & \cellcolor{green!25}↑ 3\% & \cellcolor{yellow!25}↓ 69\% & \cellcolor{yellow!25}↓ 9\% & \cellcolor{green!25}↑ 24\% \\ \hline
styleT                   & \cellcolor{green!25}$\approx$ & \cellcolor{yellow!25}↓ 2\% & \cellcolor{yellow!25}↓ 3\% & \cellcolor{green!25}$\approx$ & \cellcolor{green!25}$\approx$ & \cellcolor{green!25}$\approx$ & \cellcolor{yellow!25}↓ 3\% & \cellcolor{green!25}$\approx$ & \cellcolor{yellow!25}↓ 3\% & \cellcolor{green!25}↑ 4\% \\ \hline
\end{tabular}
}
\vspace*{0.2cm}
\end{table}

Table \ref{tab:rq4_fault} shows our fault-revealing ability measure in terms of the number of clusters in the latent space of each ADS-DNN for our $D_{\mathit{sim}}$ test sets before and after translation. As discussed in Section~\ref{sec:metrics}, our fault-revealing ability measure acts as a proxy for the number of faults a dataset can reveal in an ADS-DNN~\cite{clustering, clustering:GIST}. Note that for all the results in Table \ref{tab:rq4_fault}, we  use the HDBSCAN~\cite{hdbscan} clustering method with the same configurations.  In Table \ref{tab:rq4_fault}, we highlight cells in green (respectively, yellow) when the number of faults revealed by a $D_{\mathit{sim}}$ test set increases (respectively, decreases) after translation compared to before translation. As the table shows, except for two cases highlighted in yellow, the $D_{\mathit{sim}}$ test sets find more faults after translation than before. Interestingly, for the lane-keeping task, SAEVAE and cycleG are able to translate the $D_{\mathit{sim}}$ test sets into test sets that can find considerably more faults after translation compared to before translation. For the object-detection task, the $D_{\mathit{sim}}$ test set translated by cycleG finds the same number of faults as before translation, and the $D_{\mathit{sim}}$ test set translated by SAEVAE finds only one fewer fault than before translation.

\begin{table}[t]
\centering
\caption{Number of clusters identified in the latent space of ADS-DNNs as a proxy for fault-revealing ability of the $\text{D}_{\text{sim}}$ test sets before translation (first row), and after translation by SAEVAE (2nd row), cycleG (3rd row) and styleT (4th row)}
\label{tab:rq4_fault}
\vspace*{-0.2cm}
\scalebox{0.8}{
\begin{tabular}{|l|l|l|l|l|l|}
\hline
 & Autumn & Chauffeur & Dave2 & Epoch &  YOLO \\ \hline
$\text{D}_{\text{sim}}$  & 2      & 2         & 2     & 15 &  3   \\ \hline
SAEVAE   & \cellcolor{green!25}83     & \cellcolor{green!25}82        & \cellcolor{green!25}104   & \cellcolor{green!25}50 & \cellcolor{yellow!25} 2   \\ \hline
cycleG   & \cellcolor{green!25}98     & \cellcolor{green!25}94        & \cellcolor{green!25}82    & \cellcolor{green!25}80  & \cellcolor{green!25} 3  \\ \hline
styleT   & \cellcolor{green!25}2      & \cellcolor{green!25}2         & \cellcolor{green!25}2     & \cellcolor{yellow!25}5    & \cellcolor{green!25} 5 \\ \hline
\end{tabular}
}
\vspace*{0.3cm}
\end{table}

Table \ref{tab:rq4_blackbox} presents the black-box diversity results in terms of the GD and SD metrics, as described in Section~\ref{sec:metrics}, for our $D_{\mathit{sim}}$ test sets before and after translation. The results are calculated by subtracting the pre-translation metric values from the post-translation metric values and then dividing by the pre-translation values. As the table shows, according to the GD metric, the diversity of our $D_{\mathit{sim}}$ test sets is preserved after translation compared to before translation. While the SD metric indicates that test diversity may decline by approximately $3$\% to $33$\% after translation with styleT incurring the least decline for both ADS tasks.

\begin{table}[t]
\centering
\caption{The differences in test diversity for the $\text{D}_{\text{sim}}$ test sets before and after translation in terms of Geometric Diversity (GD) and Standard Deviation (SD): $\approx$ means the test diversity is preserved by the translator; and ↓ indicates that the translator decreases the test diversity}
\label{tab:rq4_blackbox}
\vspace*{-0.2cm}
\scalebox{0.8}{
\begin{tabular}{|l|c|c|c|c|}
\hline
\textbf{Task} & \textbf{Metric} & \textbf{SAEVAE} & \textbf{cycleG} & \textbf{styleT} \\ \hline
\multirow{2}{*}{\textbf{Lane Keeping}} & GD & \cellcolor{green!25}$\approx$ & \cellcolor{green!25}$\approx$ & \cellcolor{green!25}$\approx$ \\ \cline{2-5}
                                       & SD & \cellcolor{yellow!25}↓ 33.03\% & \cellcolor{yellow!25}↓ 24.32\% & \cellcolor{yellow!25}↓ 11.07\% \\ \hline
\multirow{2}{*}{\textbf{Object Detection}} & GD & \cellcolor{green!25}$\approx$ & \cellcolor{green!25}$\approx$ & \cellcolor{green!25}$\approx$ \\ \cline{2-5}
                                           & SD & \cellcolor{yellow!25}↓ 4.04\% & \cellcolor{yellow!25}↓ 16.11\% & \cellcolor{yellow!25}↓ 3.36\% \\ \hline
\end{tabular}

}
\vspace*{0.2cm}
\end{table}

\begin{tcolorbox}[breakable,colback=gray!10!white,colframe=black!75!black]
\textbf{RQ4:} The results obtained by five ADS-DNNs and three translators for two ADS tasks, measured using five metrics that capture the coverage, diversity, and fault-revealing ability of test data for ADS-DNNs, show that translators do not lead to a major decline in test data quality. In particular, the fault-revealing ability and the diversity, measured by GD, as well as the coverage in most cases, are preserved or improved by our best translator, SAEVAE.
\end{tcolorbox}

\subsection{RQ5: Translators' time overhead} 
\label{subsec:rq4-old}
In this research question, we evaluate the time needed to process each simulation frame both with and without the use of translators. In order to measure the average processing time per simulation frame during online testing, we introduce two concepts: \emph{simulation execution time} and \emph{simulation duration}. Specifically,  simulation execution time is concerned with how long it takes to run the simulation in real-world terms, while simulation  duration captures the amount of time that passes within the simulated environment. Execution time is typically larger than simulation duration for graphical, physics-based simulators such as ADS simulators due to heavy  mathematical computations and image rendering. For example, executing a simulation scenario capturing 10 seconds of the real world may take up to 40 seconds of computing time. We calculate the  time needed to process each simulation frame by dividing the simulation execution time with the simulation  duration.

\begin{figure}[t]
    \centering
    \includegraphics[width=0.85\columnwidth]{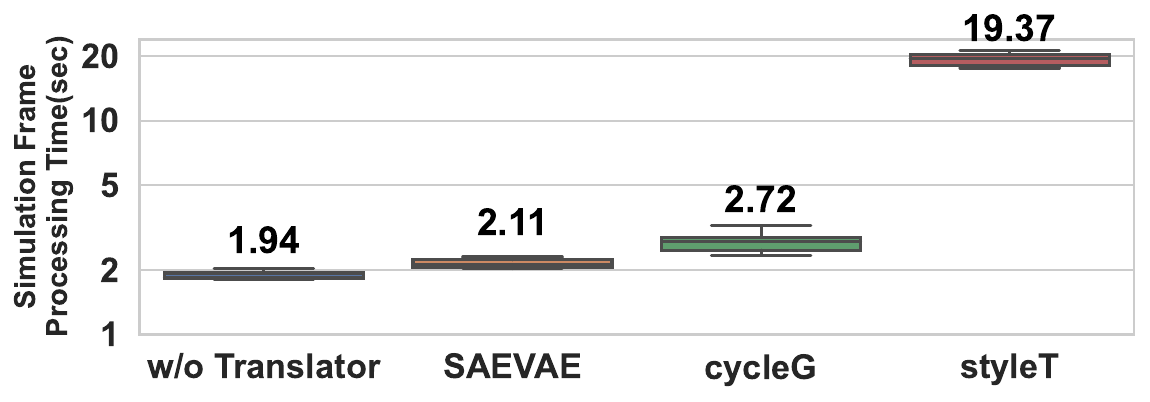}
    \caption{Simulation frame processing time in online testing without translators and using different translators}
    \label{fig:rq5_online}
\end{figure}

Figure~\ref{fig:rq5_online} shows the distributions of the time required to process each simulation frame, measured from the simulations of RQ3. The data is shown for simulations executed both without a translator and with one of the SAEVAE, cycleG, or styleT translators. 

As we expect, the lowest processing time  with a mean value of $1.94$sec is when we do not use any translator. The highest processing time belongs to styleT. SAEVAE has a mean processing time of 2.1sec, which is lower than the processing times of both cycleG and styleT, and is  a mere 0.16sec higher than the processing time observed without any translators.

\begin{tcolorbox}[breakable, colback=gray!10!white,colframe=black!75!black]
\textbf{RQ5:} The SAEVAE and cycleG translators lead to negligible time overhead during online testing. Specifically, SAEVAE leads to an average increase in simulation time of 8\%, making its usage practical for online testing.
\end{tcolorbox}

\subsection{RQ6: Online vs offline results correlation}
\label{subsec:rq5}

To answer this research question, we compute the Pearson and Spearman correlations between the MAE and OOB values, as well as between the MAE and \#OOB values, obtained from 100 simulation scenarios both with and without the use of translators. Similar to RQ3, the OOB and \#OOB values are obtained by executing each scenario four times using our four lane-keeping ADS-DNNs without any translation, and twelve more times for each pair of our ADS-DNNs with each of our three translators separately. To compute the MAE values for individual simulation images, we need a ground truth steering angle label for each such image. To obtain the ground truth labels, we designed and implemented a PID controller within the BeamNG simulator. This step was necessary because BeamNG does not provide ground truth steering angles when the vehicle maneuver is controlled by an ADS-DNN. Provided with ground truth steering angle labels, we could compute MAE values for each simulation image using our lane-keeping ADS-DNNs: first without any translation, and then separately after applying each translator.

Table~\ref{tab:rq5_corr} shows the differences in average correlation values between online and offline test results with translators compared to those without translators. Specifically, Table \ref{tab:rq6(a)} shows the differences in average correlations  between MAE and OOB, and  Table \ref{tab:rq6(b)} shows the differences in average correlations  between MAE and \#OOB. To ensure that the individual correlation values are statistically significant, we compute the p-values for all computed correlations in Table~\ref{tab:rq5_corr}. These p-values, which  are available in our replication package~\cite{github},  are zero or near zero indicating high confidence in the correlation values. All correlation values are available online~\cite{github}.

\begin{table}[t]
\caption{The differences in average correlations between online and offline test results with translators compared to those without translators: Positive values indicate that translators improve the correlation, while negative values indicate that they decrease the correlation}~\label{tab:rq5_corr}
\vspace*{-0.5cm}
\centering

\hspace{.17cm}  
\begin{subtable}[t]{0.45\textwidth}
\centering
\caption{Differences in average correlations between MAE and OOB}~\label{tab:rq6(a)}
\scalebox{0.8}{\begin{tabular}{|l|l|l|l|l|l|}
\hline
\textbf{Translator} & \textbf{Correlation} & \textbf{Autumn} & \textbf{Chauffeur} & \textbf{Dave2} & \textbf{Epoch} \\ \hline
\multirow{2}{*}{\textbf{SAEVAE}} & Pearson & -0.05 & 0.33 & 0.01 & -0.03 \\
\cline{2-6}
                                 & Spearman & 0.08 & 0.22 & 0.03 & -0.34 \\ \hline
\multirow{2}{*}{\textbf{cycleG}} & Pearson & -0.15 & 0.38 & 0.42 & -0.2 \\
\cline{2-6}
                                 & Spearman & -0.15 & 0.23 & 0.54 & -0.3 \\ \hline
\multirow{2}{*}{\textbf{styleT}} & Pearson & 0.04 & -0.09 & -0.1 & 0 \\
\cline{2-6}
                                 & Spearman & 0.11 & -0.05 & 0.07 & -0.03 \\ \hline
\end{tabular}}
\end{subtable}
\hfill
\vspace*{.2cm}
\begin{subtable}[t]{0.45\textwidth}
\caption{Differences in average correlations between MAE and \#OOB}~\label{tab:rq6(b)}
\centering
\scalebox{0.8}{\begin{tabular}{|l|l|l|l|l|l|}
\hline
\textbf{Translator}              & \textbf{Correlation} & \textbf{Autumn} & \textbf{Chauffeur} & \textbf{Dave2} & \textbf{Epoch} \\ \hline
\multirow{2}{*}{\textbf{SAEVAE}} & Pearson              & 0.87            & 1.32               & -0.10          & 0.29           \\ \cline{2-6} 
                                 & Spearman             & 0.83            & 1.18               & -0.12          & 0.48           \\ \hline
\multirow{2}{*}{\textbf{cycleG}} & Pearson              & 0.03            & -0.21              & 0.42           & -0.07          \\ \cline{2-6} 
                                 & Spearman             & 0.22            & -0.26              & 0.49           & -0.05          \\ \hline
\multirow{2}{*}{\textbf{styleT}} & Pearson              & -0.23           & -0.27              & -0.06          & 0.05           \\ \cline{2-6} 
                                 & Spearman             & 0.03            & -0.19              & -0.10          & 0.18           \\ \hline
\end{tabular}}
\end{subtable}
\end{table}

A positive difference in Table~\ref{tab:rq5_corr} indicates that translators improve the correlation between online and offline test results, while a negative difference indicates a decrease. Overall, when we use SAEVAE, the correlation between MAE and OOB increases or remains almost the same for three out of our four ADS-DNNs, namely, Autumn, Chauffeur, and Dave2. Similarly, using SAEVAE, the correlation between MAE and \#OOB increases for three out of our four ADS-DNNs, namely, Autumn, Chauffeur, and Epoch. In cases where using SAEVAE does not lead to an increase in correlation, i.e., those represented by a negative difference in Table~\ref{tab:rq5_corr}, the reduction is relatively small: around $-0.1$ for Dave2 in Table~\ref{tab:rq6(b)}, and $-0.03$ and $-0.34$, respectively, for Pearson and Spearman correlations for Epoch in Table~\ref{tab:rq6(a)}. The overall increase in the correlations between MAE and OOB, as well as between MAE and \#OOB, indicates a greater similarity between online and offline testing results when SAEVAE is used in ADS testing. This implies that when SAEVAE is used, faults identified during online testing could potentially be revealed during the less expensive offline testing.

\begin{tcolorbox}[breakable,colback=gray!10!white,colframe=black!75!black]
\textbf{RQ6:} SAEVAE improves the correlation between offline and online test results for all four of our ADS-DNNs at least based on one of the online testing metrics, OOB or \#OOB. Improving the correlation between offline and online testing helps identify faults during the less costly offline testing stage, thus reducing the overall cost of ADS testing.
\end{tcolorbox}

\subsection{Threats to Validity and Discussions}
\label{subsec:threats}

\subsubsection{Internal validity.} \label{subsec:intval}To ensure fair comparisons of the translators in our study,  we used identical datasets for training and tuning ADS-DNNs and translators, and set consistent time budgets for both training and testing. We rigorously subjected all ADS-DNNs and translators to  consistent hyperparameter tuning via  grid search. Further, we trained multiple translators and ADS-DNNs and used the best performing ones for the experiments. We fixed the test sets used to compute the VAE reconstruction errors and the MAE accuracy results for RQ1 and RQ2, respectively.

Regarding the ground-truth labels, as shown in Table~\ref{tab:train}, all except one of the labelled datasets used in our study, including those used for ADS-DNN training and validation, are public-domain datasets and pre-labelled. Specifically, we only need to generate labels for the lane-keeping synthetic images produced by BeamNG. These labels are required solely to compute the MAE results in RQ2 and RQ6 and are not necessary for any training or validation. ADS-DNN models are trained using public, pre-labeled datasets, and translators do not require labels for their training.

As discussed in Section~\ref{subsec:setupdnn}, for RQ2, we use BeamNG.AI to generate labels, a common practice for labelling synthetic images for ADS-DNN offline testing~\cite{offline_online}. To ensure the generated labels are accurate, we set the \texttt{drive\_in\_lane} flag of BeamNG.AI to true, ensuring that it only drives within the lanes it can legally drive in~\cite{beamngai}. For RQ6, as discussed in Sections~\ref{subsec:setupdnn} and~\ref{subsec:rq5}, due to technical issues with BeamNG, we developed a separate PID controller to generate the ground-truth steering angle labels. We integrated the controller into a closed-loop with the BeamNG simulator and tuned it interactively using BeamNG's feedback, ensuring that the controller generates steering angle values to maneuver the ego vehicle. The controller error is  the distance between the ego vehicle and the center of the lane. Following standard control theory guidelines~\cite{control_dorf}, we calibrated the P, I, and D coefficients. We then tested the controller using several simulations to ensure its accuracy. The PID controller is available in our replication package~\cite{github}. Finally, in both RQ2 and RQ6, we evaluate all the translators and ADS-DNNs using the same automatically generated labels, thus not favoring any particular method.

Modifying images using generative deep-learning models may lead to invalid input images~\cite{when_and_why}. In our context, an image is deemed invalid if it lacks roads or vehicles, or if the road and vehicles are so malformed that they are unrecognizable to humans, preventing the assignment of a steering angle or the detection of different vehicles in the image. The translators we used in our study can potentially transform ADS images into invalid images, which may cause inaccuracies in our experiments. To avoid this issue, a third party  manually inspected  the images generated by the translators in our study for RQ1 and RQ2; all the images were deemed valid.

\subsubsection{External validity.} \label{subsec:extval} Our experiments use  five different ADS-DNNs, three different translator approaches, and public datasets for two different ADS tasks. All these ADS-DNNs are community-contributed models and are extensively used as benchmarks for ADS testing~\cite{offline_online,comparing_offline_online,stoccoicst,deeptest,surprise}. Two of them, in addition to appearing in research papers, have been used by industry: YOLO has been extensively used by NVIDIA to showcase its GPU performance for ADS applications~\cite{yolonvidia, digits, tensorrt}, and Dave2 has been successfully employed in real-world road testing conducted by NVIDIA~\cite{dave2roadtest}.  For online testing, we use BeamNG, a leading open-source simulator widely referenced for virtual and hybrid testing in ADS research, and used by the software testing community for benchmarking and competitions.~\cite{wherearewe,sbftgithub}. We evaluate the impact of translators on test data quality using five different metrics from the literature that serve as proxies for coverage, diversity, and the fault-revealing ability of test data for DNNs. To further demonstrate generalizability,  our study can be replicated for other ADS-DNNs, such as  Apollo~\cite{apollo} and Autoware~\cite{autoware}, other simulators, such as CARLA~\cite{carlapaper}, and other test data quality metrics.

\section{Related Work}
\label{sec:related}

The gap between real-world and synthetic images is a widely acknowledged problem in robotics and ADS~\cite{gap_sim2real, afsoon-robotics, offline_online, lei}. According to a recent survey of industry professionals in robotics, over half of engineers and practitioners cite lack of realism and accuracy  as the primary reason for not using simulators in their projects~\cite{afsoon-robotics}. The discrepancies between simulator-generated and real-world datasets particularly weakens the effectiveness of ADS testing, leading to overly negative and misrepresented assessments of their capabilities when tested using simulators~\cite{offline_online,mind_the_gap}. In addition to discrepancies between real-world and simulator-generated images, recent studies have reported variations in testing results from different simulators, which may partly stem from the misalignment of images produced by these various simulators~\cite{digital_twins, two_is_better}. Our evaluation results confirm the gap between real-world and simulator-generated images, based on public-domain datasets and those obtained from a well-established online testing setup. Further,  we  demonstrate that this gap can significantly impact ADS testing results.

Several approaches for image-to-image translation are proposed in the deep-learning community to support different use cases~\cite{pix2pix, cyclegan, dclgan, gatys}. For example,  Pix2Pix~\cite{pix2pix} has been used to transform grayscale images to RGB images after training on paired datasets, while CycleGAN~\cite{cyclegan} and DCLGAN~\cite{dclgan}  have been used for converting images between summer and winter scenes without the need for paired datasets in their training. In the software testing literature, deep-learning image-to-image translators have been used to generate test inputs representing various weather conditions~\cite{deeproad} or to conduct adversarial attacks~\cite{deepbillboard}. 
We use  image-to-image translators suitable for mitigating the gap between simulator-generated and real-world images in ADS testing, which do not require paired datasets for training.  We demonstrate that image-to-image translators preserve the quality of simulator-generated test images in terms of coverage, diversity, and fault-revealing ability while reducing the dissimilarities between the distributions of simulator-generated test images  and real-world training images. Therefore, translators enable the accurate evaluation of ADS-DNNs using synthetic test images.

To ensure the quality of ADS-DNNs, multiple levels of testing are required, including testing with static images and testing in an online, closed-loop mode, using simulators or real cars~\cite{zhang_ml_testing}. Recent studies have compared various testing levels for ADS-DNNs, such as simulator versus real-world offline dataset results~~\cite{offline_online,comparing_offline_online,codevilla2018offline}, microscale ADS testing platforms, i.e., donkey cars, versus simulators~\cite{system_level,mind_the_gap}, and multiple simulators~\cite{digital_twins,two_is_better}. While all studies indicate mismatches in results obtained from different testing levels, thus questioning the reliability of ADS-DNN testing using simualtors, only a few have attempted to mitigate these mismatches. In particular, Stocco et al.~\cite{system_level,mind_the_gap, stoccoicst} and Biagiola et al.~\cite{two_is_better} use CycleGAN to narrow the gap in image distributions between simulator outputs and a donkey-car setup, and between BeamNG and Udacity simulators, respectively. While existing research investigates the impact of translators in reducing the gap between different image distributions in offline testing, we additionally evaluate their effectiveness in reducing failures in online testing. To ensure that translators can be reliably integrated into our ADS testing practices, we demonstrate that they do not compromise important test-data qualities through metrics such as test coverage, diversity, and fault-revealing ability.  Further, we demonstrate that our proposed image-to-image translation approach, SAEVAE, can be efficiently integrated into online testing. Finally, we show that  translators can increase the correlation between offline and online testing results.

\section{Conclusion}
\label{sec:conclusion}

We studied the effectiveness, reliability, and efficiency of using image-to-image translators for testing Autonomous Driving Systems (ADS). We demonstrated the effectiveness of these translators in bridging the gap between real-world and synthetic images for ADS testing, as well as improving accuracy for both offline and online ADS testing. We showed that translators can preserve the coverage, diversity, and fault-revealing capabilities of synthetic images for ADS-DNN testing. Finally, we showed that SAEVAE and CycleGAN translators incur negligible time overhead during online testing and can potentially reduce the need for online testing by increasing the correlation between online and offline testing.

Our work contributes a public dataset~\cite{github} of \linebreak simulator-generated images for ADS lane keeping  corresponding to the Udacity Jungle dataset. Our dataset along with  KITTI and vKITTI can be used  in future research and experimentation to further improve the accuracy gap for ADS testing.  We plan to enhance our work by developing online test setups for other ADS tasks and experimenting with alternative ADS-DNNs and ADS datasets. 

Our complete \textbf{replication package}  is available online~\citep{github}.

\vspace*{.5em}\textbf{Acknowledgements.} We gratefully acknowledge the financial support from NSERC of Canada through the Discovery program.


\newpage 

\onecolumn \begin{multicols}{2}

\bibliographystyle{unsrt}
\bibliography{bib/ref}
\end{multicols}



\end{document}